\documentclass{article}

\usepackage{amsmath,amsfonts,amssymb,amsthm}
\usepackage{setspace,graphicx}
\usepackage{natbib}
\usepackage{bm}
\usepackage{ifthen}
\usepackage{verbatim}
\usepackage{color}
\usepackage{dsfont}
\usepackage{tikz}
\usepackage{setspace,graphicx}
\usepackage{sgame}
\usepackage{bm}
\usepackage{multirow}
\usepackage{ifthen}
\usepackage{verbatim}
\usepackage{xcolor}
\usetikzlibrary{arrows.meta}
\usepackage{xcolor}

\usepackage{listings}
\UseRawInputEncoding

\theoremstyle{plain}
\theoremstyle{definition}
  \newtheorem{theorem}{Theorem}

  \newtheorem{definition}{Definition}
  
  \newtheorem{example}{Example}
  \newtheorem{lemma}{Lemma}
  
  \newtheorem{proposition}{Proposition}
  \newtheorem{remark}{Remark}
  \theoremstyle{remark}

\def\EE{\mathbb{E}}
\def\RR{\mathbb{R}}

\begin{document}


\title{Anonymous voting in a heterogeneous society}

\author{
  Yaron Azrieli\thanks{Department of Economics, The Ohio State University,  
  1945 North High Street, Columbus, OH 43210, USA.  
  Email: \texttt{azrieli.2@osu.edu}} 
  \and
  Ritesh Jain\thanks{University of Liverpool Management School.  
  Email: \texttt{ritesh.jain@liverpool.ac.uk}} 
  \and
  Semin Kim\thanks{School of Economics, Yonsei University,  
  50 Yonsei-ro, Seodaemun-gu, Seoul 03722, Republic of Korea.  
  Email: \texttt{seminkim@yonsei.ac.kr}}
}

\maketitle

\begin{abstract}
We study the design of voting mechanisms in a binary social choice environment where agents' cardinal valuations are independent but not necessarily identically distributed. The mechanism must be anonymous -- the outcome is invariant to permutations of the reported values. We show that if there are two agents then expected welfare is always maximized by an ordinal majority rule, but with three or more agents there are environments in which cardinal mechanisms that take into account preference intensities outperform any ordinal mechanism.




\end{abstract}

\newpage
\section{Introduction}\label{sec:introduction}

Imagine a Homeowners Association (HOA) whose members occasionally need to decide whether to undertake capital projects. Any such decision may impact members differently: Renovating the pool may be extremely important for member A, only slightly useful for member B, and definitely not worth it for member C. Furthermore, it is conceivable that member A usually values such projects much more than member B, so these two members are different even from the point of view of an outside observer who does not know their valuations. What should be the voting rule the HOA uses to make decisions, provided that ethical considerations and regulation forces it to treat all members symmetrically and prohibit side transfers?

A straightforward solution is to use some form of majority rule, perhaps with approval threshold different than 50\%. Such rules are \emph{anonymous} in the sense that the outcome is invariant to permutations of the votes, i.e., the identity of voters need not be known to figure out the outcome. Such rules are also \emph{ordinal} -- they do not take into account the intensity of members' preferences, only whether they are for or against the project. In this paper we ask whether a social planner can improve members' total expected welfare by designing a cardinal mechanism that does take into account preference intensities. The mechanism must be anonymous, i.e., invariant to permutations of the reported intensities; and it must be incentive compatible -- truthful reporting is optimal when everyone else is truthful. We show that if there are only two agents then majority rules cannot be improved upon by cardinal mechanisms (Theorem \ref{theorem-n=2}). Surprisingly, if there are at least three agents then, depending on the environment, cardinal mechanisms may yield higher welfare than any ordinal mechanism (Theorem \ref{theorem-n>3}).

More formally, we consider a standard Bayesian mechanism design setup with $n$ agents who face a binary choice between keeping the status-quo or implementing a reform. Valuations are independently drawn, but the distributions may differ across agents (the support is the same). We focus on direct mechanisms in which each agent reports their value and the mechanism outputs the probability with which the reform is implemented. We require the mechanism to be Bayesian Incentive Compatible (BIC) and anonymous. 

Notice that in this setup the BIC constraints are extremely demanding: The expected probability that the reform is chosen must be constant in an agent's report, so long as this report indicates the same ordinal preference between the alternatives. Thus, on average, the intensity of an agent's preferences can't affect the probability with which the reform is chosen. And indeed, without anonymity constraints, ordinal mechanisms are always optimal \citep{AzrieliKim2014}. But once anonymity is imposed, ordinal mechanisms become overly restrictive, forcing agents' influence on the outcome to depend only on the probabilities with which they prefer each alternative. Cardinal mechanisms allow flexibility in giving greater influence to agents that have greater stakes on average, even subject to the BIC and anonymity constraints. The proof of Theorem \ref{theorem-n>3} follows this intuition and constructs a class of examples where cardinal mechanisms outperform ordinal mechanisms.

\section{Set up}\label{sec:model}

We consider a group of $n\ge 2$ individuals who needs to choose between implementing a Reform or keeping the Status quo. Let $N=\{1,\ldots,n\}$ be the set of agents and $\{R,S\}$ the set of alternatives. We normalize the utilities so that all agents receive utility of zero if $S$ is chosen. If $R$ is implemented then the utility of agent $i$ is $v_i\in \RR$. We refer to $v_i$ as $i$'s value and assume that it is $i$'s private information. All agents, as well as the social planner, share a common prior belief regarding the distribution of values. Namely, the distribution of $v_i$ is $G_i$, and the values of different agents are statistically independent. It will be convenient to let $\tilde{v}=\left(\tilde{v}_1,\ldots,\tilde{v}_n\right)$ be independent random variables such that $\tilde{v}_i$ is distributed according to $G_i$, $i=1,\ldots, n$.

Throughout the paper we make the following assumptions:

\begin{enumerate}
\item All the $\tilde{v}_i$'s have the same support, denoted $V$, and $0\notin V$.

\item Each $\tilde{v}_i$ has a finite expectation: $\EE \left(\left|\tilde{v}_i\right|\right)<+\infty$.

\item Each $\tilde{v}_i$ attains both positive and negative values with positive probability: $p_i:= 1-G_i(0)\in (0,1)$.
\end{enumerate}

Thus, agents are never indifferent and they may prefer either alternative. We emphasize that values of different agents need not be identically distributed; indeed, we are particularly interested in cases with ex-ante heterogeneity.

A Social Choice Function (SCF) is a Borel measurable mapping $f:V^n\to [0,1]$ that determines the probability that $R$ is implemented for every profile of values. We are interested in SCFs that satisfy anonymity and incentive compatibility properties. To define anonymity, whenever $\pi:N\to N$ is a permutation of agents' names and $v\in V^n$ is a profile of values we let $\pi v=\left(v_{\pi(1)}, \ldots, v_{\pi(n)}\right)$ be the corresponding permuted profile of values.

\begin{definition}\label{def-anonymity}
A SCF $f$ is anonymous if for every $v\in V^n$ and every permutation of agents $\pi$ it holds that
\begin{equation}\label{eqn-anonymity}
f(v)=f(\pi v).
\end{equation}
\end{definition}

Next, the definition of incentive compatibility is standard and requires that truthful reporting of values is a Bayes-Nash equilibrium of the game induced by $f$.

\begin{definition}\label{def-IC}
A SCF $f$ is Bayesian Incentive Compatible (BIC) if for every agent $i$ and every $v_i,v'_i\in V$ it holds that\footnote{Throughout we use the standard notation where a subscript $-i$ indicates that the $i$th coordinate of a vector is omitted. Expectations are calculated relative to the variables with tilde above them.}
\begin{equation}\label{eqn-IC}
v_i \EE(f(v_i,\tilde{v}_{-i})) \ge  v_i \EE(f(v'_i,\tilde{v}_{-i})).
\end{equation}
\end{definition}

The following characterization of BIC SCFs is immediate, so we omit the proof. Essentially the same result appears in \citet[Lemma 3]{SchmitzTroger2012}.

\begin{lemma}\label{lemma-IC}
A SCF $f$ is BIC if and only if for every agent $i$ there are two numbers $0\le c_i^- \le c_i^+\le 1$ such that $\EE(f(v_i,\tilde{v}_{-i}))=c_i^-$ for every $v_i<0$ and $\EE(f(v_i,\tilde{v}_{-i}))=c_i^+$ for every $v_i>0$.
\end{lemma}

The expected welfare generated by a SCF $f$ is denoted by
$$W(f)=\EE\left( f(\tilde v) \sum_{i=1}^n \tilde{v}_i\right).$$
Note that for BIC SCFs we can use Lemma \ref{lemma-IC} to rewrite $W$ as a function of $c_i^-$ and $c_i^+$ only. Indeed, for each agent $i$ let $U_i^+=\EE \left(\tilde{v}_i ~|~ \tilde{v}_i >0\right)$ and $U_i^-=\EE \left(|\tilde{v}_i| ~|~ \tilde{v}_i <0 \right)$ (both are well-defined, positive, and finite given our assumptions). Then for every BIC $f$ we have
\begin{equation}\label{eqn-welfare}
W(f) = \sum_{i=1}^n \EE\left( \tilde{v}_i f(\tilde v) \right) = \sum_{i=1}^n \int_V v_i \EE(f(v_i,\tilde{v}_{-i})) dG_i(v_i) = \sum_{i=1}^n \left[c_i^+(f) p_i U_i^+ - c_i^-(f) (1-p_i)U_i^-\right],
\end{equation}
where we write $c_i^+(f),c_i^-(f)$ rather than $c_i^+,c_i^-$ to emphasize the dependence on $f$.

Finally, an important part in the analysis is played by ordinal SCFs, i.e.\ SCFs that do not depend on the intensity of preferences. Formally, let $\chi(v) = \{i\in N ~:~ v_i>0\}$ be the coalition of agents who prefer $R$ over $S$ given value profile $v$.
\begin{definition}\label{def-ordinal}
A SCF $f$ is ordinal if $f(v)=f(v')$ whenever $\chi(v)=\chi(v')$.
\end{definition}
Note that if $f$ is both ordinal and anonymous then it depends only on the number of agents that prefer $R$ and not on their identity. In particular, for any $k\in\{1,\ldots,n\}$ we denote by $f^{(k)}$ the \emph{qualified majority rule} with threshold $k$:
\begin{equation*}
f^{(k)}(v) =
\left\{ \begin{array}{ll}
1 & \textit{if }~ |\chi(v)| \ge k \\
0 & \textit{if }~ |\chi(v)| < k.
\end{array} \right.
\end{equation*}

\section{Optimal mechanisms}

We are interested in properties of the solution of the following program:

\begin{equation}\tag{OPT}\label{eqn-max-welfare}
\max_{f}~ W(f)
\end{equation}
$$\textit{subject to } (\ref{eqn-anonymity}) \textit{ and } (\ref{eqn-IC}).$$

\subsection{Symmetric environments}\label{subsec-symmetric}
Suppose first that agents are ex-ante identical, i.e., $G_1=\ldots = G_n$. To characterize the solution to (\ref{eqn-max-welfare}) in this case, for any SCF $f$ and a coalition $T\subseteq N$ define by $\phi_f(T) = \EE\left(f(\tilde v) ~|~ \chi(\tilde v)=T\right)$ the expected probability that $R$ is chosen conditional on the event that the members of $T$ are exactly those who prefer $R$. We can then associate with any $f$ the ordinal SCF $\hat f$ defined by $\hat f(v)=\phi_f(\chi(v))$. It is not hard to show (see \cite[Lemma 2]{AzrieliKim2014} or \cite[Lemma 14]{SchmitzTroger2006}) that if $f$ is BIC then so is $\hat f$ and that $c_i^+(\hat f)=c_i^+(f)$ and $c_i^-(\hat f)= c_i^-(f)$, which by (\ref{eqn-welfare}) implies $W(f)=W( \hat f)$. Moreover, if $f$ is anonymous then the symmetry of the environment implies that $\hat f$ is anonymous as well.\footnote{See Example \ref{example-non-anonymous} below for an illustration of how to obtain $\hat f$ from $f$ and why symmetry of the environment is necessary to conclude that $\hat f$ is anonymous whenever $f$ is.}

It follows that there is an ordinal solution to (\ref{eqn-max-welfare}). Conditional on the event $\{|\chi(\tilde v)|=k\}$, if $k U_1^+ > (n-k) U_1^-$ then social welfare is maximized when $R$ is chosen, and if the reverse inequality holds then social welfare is maximized when $S$ is chosen. The following proposition summarizes the above discussion; slightly different versions of this result appear in \cite[Proposition 2]{SchmitzTroger2012} and \cite[Theorem 1]{AzrieliKim2014}.

\begin{proposition}\label{prop-symmetric}
If $G_1=\ldots = G_n$ then the qualified majority rule $f^{(\bar k)}$ solves (\ref{eqn-max-welfare}), where $\bar k=\min\{k\in\{1,\ldots,n\} : k>\frac{U_1^-}{U_1^+ + U_1^-} n\}$. In particular, the maximal expected welfare can be achieved by an ordinal SCF.
\end{proposition}

\begin{remark}
If $\frac{U_1^-}{U_1^+ + U_1^-} n$ happens to be an integer $k$ then any tie-breaking rule can be used when $|\chi(v)| = k$ without affecting the objective; otherwise, the solution is unique (up to zero probability events).
\end{remark}

\begin{remark}
In symmetric environments the anonymity constraints (\ref{eqn-anonymity}) are not binding: There exists a solution to the relaxed program without anonymity constraints that satisfies anonymity. A direct argument is as follows: If $f$ is any BIC SCF and $\pi$ is any permutation of agents' names then the SCF $\pi f$ defined by $\pi f(v) = f(\pi v)$ is also BIC and moreover $W(f)=W(\pi f)$. Thus, if $f$ is BIC and maximizes expected welfare then so is the average over all of its permutations  $\frac{1}{n!}\sum_\pi \pi f$, which satisfies anonymity.

\end{remark}

\subsection{Asymmetric environments: Two agents}\label{subsec:n=2}
Once we move away from symmetric environments, the argument underlying Proposition \ref{prop-symmetric} is no longer valid. The problem is that for BIC and anonymous $f$, the associated SCF $\hat f$ is not necessarily anonymous, so we can't conclude that restricting attention to ordinal functions is without loss. The following example illustrates this.

\begin{example}\label{example-non-anonymous}
Let $n=2$ and $V= \{-2,-1, 1, 2\}$. The distributions of values for agent 1 (rows) and agent 2 (columns) are given in the following table, along with a particular SCF $f$:

\begin{center}
\begin{tabular}{c|cc|cc|}

          & (1/2) & (1/4) & (1/8) & (1/8) \\
      & -2 & -1 & 1 & 2          \\ \hline
      (1/6)   2 & 0 & 1 & 1 & 1 \\
      (1/6)   1 & 0 & 1 & 1 & 1 \\ \hline
      (1/6) -1 & 0 & 1 & 1 & 1 \\
      (1/2) -2 & 1 & 0 & 0 & 0 \\ \hline
\end{tabular}
\end{center}
\medskip

It is immediate to verify that $f$ is BIC and anonymous. The associated `ordinal conditional expectation' $\hat{f}$ is given by:

\begin{center}
\begin{tabular}{c|cc|cc|}

        & (1/2) & (1/4) & (1/8) & (1/8) \\
       & -2 & -1 & 1 & 2          \\ \hline
      (1/6)   2 & 1/3 & 1/3 & 1 & 1 \\
      (1/6)   1 & 1/3 & 1/3 & 1 & 1 \\ \hline
      (1/6)  -1 & 7/12 & 7/12 & 1/4 & 1/4 \\
      (1/2)  -2 & 7/12 & 7/12 & 1/4 & 1/4 \\ \hline
\end{tabular}
\end{center}
\medskip
Thus, while $\hat f$ is BIC and maintains the same expected welfare as $f$, it is not anonymous.
\end{example}

Despite this, it turns out that in the case of two agents we can restrict attention to qualified majority rules without loss.

\begin{theorem}\label{theorem-n=2}
For every pair of distributions $G_1,G_2$ (at least) one of the qualified majority rules $f^{(1)}$ or $f^{(2)}$ solves (\ref{eqn-max-welfare}). In particular, the maximal expected welfare can be achieved by an ordinal SCF.
\end{theorem}

The proof of Theorem \ref{theorem-n=2} is in Appendix \ref{subsec-proof1}. The key is to obtain bounds on the impact that individual agents can have on the outcome of the mechanism. Specifically, we prove that for any BIC and  anonymous SCF $f$ the inequalities
$$p_1c_2^+(f) - (1-p_1)c_2^-(f) \le p_1^2$$
and
$$p_2c_1^+(f) - (1-p_2)c_1^-(f) \le p_2^2$$
hold (recall that $p_i=1-G_i(0)$ is the probability that $i$ prefers $R$ over $S$). We then argue that in any solution to (\ref{eqn-max-welfare}) these two inequalities must hold as equalities, from which the result follows.

\subsection{Asymmetric environments: More than two agents}\label{subsec:n>2}

Surprisingly, when there are more than two agents it is sometimes possible to use the reported intensity of preferences to improve welfare.

\begin{theorem}\label{theorem-n>3}
For every $n\ge 3$ there exist distributions $G_1,\ldots,G_n$ for which no ordinal SCF solves (\ref{eqn-max-welfare}).
\end{theorem}

The proof of Theorem \ref{theorem-n>3} is in Appendix \ref{subsec-proof2}. We sketch here the argument for the case of $n=3$ agents. Let $V=\{-100,-1,1,10\}$. We consider an environment with two types of agents: Agents 1 and 2 are `high-stakes' agents in the sense that their values are very likely $10$ or $-100$, and agent 3 is `low-stake' -- her value is very likely $1$ or $-1$. More precisely, suppose that for some small $\epsilon>0$ the value distributions are as in the following table.

\begin{center}
\begin{tabular}{c||c|c|c|c|}
$V$ & -100 & -1 & 1 & 10 \\
\hline
$G_1$ and $G_2$ & 0.5-$\epsilon$ & $\epsilon$ & $\epsilon$ & 0.5-$\epsilon$   \\ 
\hline
$G_3$ &  $\epsilon$ & 0.5-$\epsilon$ & 0.5-$\epsilon$ & $\epsilon$ \\
\hline   
\end{tabular}
\end{center}

\underline{Step 1:} Best ordinal rule.\\
Restricting attention to ordinal (and anonymous) rules, for all sufficiently small $\epsilon$ the optimal SCF is the unanimity rule $f^{(3)}$, i.e., $R$ is implemented if all three agents have positive values and $S$ is chosen otherwise. Indeed, if only two out of the three agents prefer $R$ then there is almost $2/3$ chance that the remaining agent has a value of -100, so it is better to stick with $S$. It follows that the welfare under the best ordinal anonymous rule converges to $\left(\frac{1}{2}\right)^3[10+10+1]=\frac{21}{8}$ as $\epsilon\to0$.

\medskip

\underline{Step 2:} A better cardinal rule when $\epsilon=0$.\\
It would be useful to consider the limiting case $\epsilon=0$, even though it violates our assumption that all distributions have the same support. Specifically, we describe an anonymous cardinal rule $f^*$ that when $\epsilon=0$ is BIC and yields higher welfare than $\frac{21}{8}$. We emphasize that the domain of $f^*$ is the entire $V^3$, i.e., agents are allowed to report values that have zero probability, and the BIC constraints must hold for such reports as well. The following table specifies $f^*$ and compares it to the optimal ordinal rule $f^{(3)}$.\footnote{We use multiset notation for the profiles of values to remind the reader that anonymity forces invariance to permutations of values.}

\begin{center}
\begin{tabular}{c||c|c|c|c|c|}
Profile of reports & all positive & $\{10,10,-1\}$ & $\{1,1,-100\}$ & $\{-100,-100,-100\}$ & otherwise \\
\hline
$f^*$ & 1 & 1 & 1 & 1 & 0  \\ 
\hline
$f^{(3)}$ & 1 & 0 & 0 & 0 & 0  \\ 
\hline 
\end{tabular}
\end{center}

Notice that $f^*$ is identical to $f^{(3)}$ except for three value profiles, two of which have zero probability. Thus, for the purpose of comparing welfare, the only difference is at the profile $\{10,10,-1\}$, where $f^*=1$ and $f^{(3)}=0$. It follows that the welfare under $f^*$ is $\frac{21}{8}+\frac{19}{8}=5$, strictly greater than under $f^{(3)}$.

Checking that $f^*$ is BIC is tedious but straightforward: If agent 1 reports either a value of -100 or -1 then the probability that $R$ is selected is 0, and if she reports a value of either 10 or 1 then this probability is 0.5; clearly, the same is true for agent 2; for agent 3, whatever value they report the probability of $R$ is 0.25. This proves that $f^*$ is BIC and moreover demonstrates that, despite it being anonymous, the high-stakes agents 1 and 2 have greater influence on the outcome than agent 3.  

\medskip

\underline{Step 3:} Approximating $f^*$ when $\epsilon\to 0$.\\
The last step is to show that for any $\epsilon>0$ sufficiently small we can find a BIC and anonymous SCF $f_\epsilon$ such that $\lim_{\epsilon\to 0} f_\epsilon= f^*$. Indeed, since the welfare $W$ is continuous in the value distributions and in the SCF, that would imply that the welfare of $f_\epsilon$ in the $\epsilon$ environment converges to 5 (the welfare of $f^*$ in the $\epsilon=0$ environment). On the other hand, from step 1 we know that the welfare of the best ordinal rule $f^{(3)}$ converges to $\frac{21}{8}$, so that $f_\epsilon$ yields higher welfare than any ordinal anonymous rule whenever $\epsilon$ is sufficiently small. 

Explicitly constructing SCFs $\{f_\epsilon\}$ with these properties is not easy, so instead we use known results from linear algebra to prove that the correspondence from $\epsilon$ to the set of BIC and anonymous SCFs is lower hemi-continuous at $\epsilon=0$. Namely, we show that the rank of the matrix that describes the BIC equality constraints and anonymity constraints is constant for $\epsilon\in[0,\bar\epsilon]$. This implies that the linear subspace of solutions is continuous at $\epsilon=0$; the other inequality constraints can be handled directly to establish lower hemi-continuity.

\section{Remarks and related literature}

The analysis of the welfare generated by voting rules in binary choice environments has a long history. Early works include \citet{Rae1969, Badger72} and \citet{Curtis72}, while \citet{SchmitzTroger2012} provides a more modern treatment that allows for cardinal utilities but restricts attention to symmetric environments. In asymmetric environments, several papers \citep{BarberaJackson2006, Fleurbaey2008, AzrieliKim2014} show the connection between welfare maximization and weighted majority rules. 

A growing literature looks at the implications of imposing symmetry constraints in mechanism design environments that may be asymmetric. For example, \citet{deb2017discrimination} and \citet{chen2025optimal} consider the design of revenue maximizing auctions, \cite{korpela2018procedurally} and \cite{barlo2023anonymous} analyze complete-information Nash implementability, and \cite{azrieli2018symmetric} prove a revelation principle type result in an abstract Bayesian setup. In an ordinal binary voting environment, \cite{azrieli2018price} calculates the `price of one-person-one-vote', defined as the ratio between the welfare achievable with and without anonymity constraints. In the context of the current paper, one can consider the relaxation of program (\ref{eqn-max-welfare}) without the anonymity constraints for which the solution is always a weighted majority rule \citep{AzrieliKim2014}, and compare it to the solution of (\ref{eqn-max-welfare}), as well as to the welfare under the optimal qualified majority rule. It would be interesting to know how much of the welfare loss due to anonymity is driven by a restriction to ordinal mechanisms: In the example of the previous section it turns out that the welfare under the optimal anonymous cardinal rule is almost the same as under the optimal (non-anonymous) mechanism.\footnote{In Appendix \ref{sec-code} we provide a Python code that allows the user to enter the distributions of agents' valuations and uses the Z3 solver to calculate all three welfare values: With anonymity and ordinality constraints, with anonymity constraints only, and without any of these constraints.} 

Relatedly, one can look at the welfare loss due to a restriction to ordinal rules. In the example of the previous section, the ratio between the welfare under the optimal cardinal and ordinal rules can be made arbitrarily close to two by appropriately scaling up the values in $V$; that is, expected welfare almost doubles if cardinal rules are allowed. We were not able to find examples with higher ratios when there are three agents, but with more agents the ratio can be higher. When there are more than two alternatives, it is well-known that mechanisms that use cardinal information can increase welfare even when transfers are infeasible, see for example \cite{BorgersPostl2009}, \cite{Miralles2012}, and \cite{KIM2017}. However, \cite{EhlersMajumdar2020} prove that adding a continuity requirement to the BIC constraints forces the mechanism to be ordinal. 

The mechanisms we consider make strong assumptions regarding the planner and agents' knowledge of the environment. Specifically, the BIC constraints depend in a delicate way on the distributions of values, and this distribution is assumed to be common knowledge. Strengthening the incentive constraints to require strategy-proofness would eliminate the possibility to use cardinal mechanisms. 


Lastly, we restricted attention to a utilitarian planner whose objective is to maximize the sum of agents' expected welfare. One can instead look at other welfare criteria such as Pareto efficiency. Weighting agents' welfare differently adds another layer of heterogeneity to the problem that can either amplify or counteract the distribution heterogeneity. 


\bibliographystyle{plainnat}
\bibliography{sym_voting}


\appendix

\section{Proofs}\label{sec-proofs}

\subsection{Proof of Theorem \ref{theorem-n=2}}\label{subsec-proof1}

The proof of the theorem is based on the following two lemmas.
\begin{lemma}\label{lemma-n=2}
Fix $G_1,G_2$ and suppose $f$ is BIC and anonymous. Then
\begin{equation}\label{eqn-bound1-n=2}
p_1c_2^+(f) - (1-p_1)c_2^-(f) \le p_1^2
\end{equation}
and
\begin{equation}\label{eqn-bound2-n=2}
p_2c_1^+(f) - (1-p_2)c_1^-(f) \le p_2^2.
\end{equation}
\end{lemma}

\smallskip

\noindent \textbf{Proof of Lemma \ref{lemma-n=2}:}\\
Let $X,Y$ be i.i.d.\ random variables, each distributed according to $G_1$. Since $f$ is BIC it follows from Lemma \ref{lemma-IC} that
$$\EE\left[f(X,Y)\mathds{1}_{\{Y>0\}}\right] = \int_0^\infty \EE\left[f(X,y)\right]dG_1(y) = p_1c_2^+(f),$$
and
$$\EE\left[f(X,Y)\mathds{1}_{\{Y<0\}}\right] =  \int_{-\infty}^0 \EE\left[f(X,y)\right]dG_1(y) = (1-p_1)c_2^-(f).$$
Now, since $f$ is anonymous,
$$\EE\left[f(X,Y)\mathds{1}_{\{Y>0\}}\right] = \EE\left[f(X,Y)\mathds{1}_{\{X>0\}}\right] ~~\textit{ and }~~ \EE\left[f(X,Y)\mathds{1}_{\{Y<0\}}\right] = \EE\left[f(X,Y)\mathds{1}_{\{X<0\}}\right].$$
Combining these equalities gives
\begin{eqnarray*}
& & 2 \big(p_1c_2^+(f) - (1-p_1)c_2^-(f)\big) = \\
& & \EE\left[f(X,Y)\mathds{1}_{\{Y>0\}}\right] +  \EE\left[f(X,Y)\mathds{1}_{\{X>0\}}\right] - \EE\left[f(X,Y)\mathds{1}_{\{Y<0\}}\right] - \EE\left[f(X,Y)\mathds{1}_{\{X<0\}}\right] =\\
& & \EE\left[f(X,Y)\left(\mathds{1}_{\{Y>0\}}  + \mathds{1}_{\{X>0\}} - \mathds{1}_{\{Y<0\}} - \mathds{1}_{\{X<0\}} \right)\right] = \EE\left[f(X,Y)\left(2\cdot \mathds{1}_{\{X>0, Y>0\}} - 2\cdot \mathds{1}_{\{X<0, Y<0\}} \right) \right] =\\
& & 2\left(\EE\left[f(X,Y)\mathds{1}_{\{X>0, Y>0\}}\right] - \EE\left[f(X,Y)\mathds{1}_{\{X<0, Y<0\}}\right] \right) \le 2 \left(\EE\left[\mathds{1}_{\{X>0, Y>0\}}\right] - 0 \right) = 2p_1^2,
\end{eqnarray*}
where the inequality follows from $0\le f \le 1$. This proves (\ref{eqn-bound1-n=2}). The proof of (\ref{eqn-bound2-n=2}) is similar.
\hfill \qed

\bigskip

\begin{lemma}\label{lemma-aux}
Consider the following maximization problem, referred to as program (AUX):
\begin{eqnarray}
\max_{c=\left( c_1^+, c_1^-, c_2^+, c_2^- \right)}  W'(c) = p_1 U_1^+c_1^+ - (1-p_1) U_1^-c_1^- &+& p_2 U_2^+c_2^+ - (1-p_2) U_2^-c_2^- \nonumber\\
\label{eqn-bound1'-n=2} \textit{s.t. }~~~ p_1c_2^+ - (1-p_1)c_2^- &\le& p_1^2\\
\label{eqn-bound2'-n=2} p_2c_1^+ - (1-p_2)c_1^- &\le& p_2^2\\
\label{eqn-n=2-equality} p_1c_1^+ + (1-p_1)c_1^- &=& p_2c_2^+ +(1-p_2)c_2^-\\
\label{eqn-0-1} 0\le c_1^+, c_1^-, c_2^+, c_2^- &\le& 1.
\end{eqnarray}
Then either $c^{(1)}=(c_1^+=c_2^+=1,~ c_1^-=p_2,~ c_2^-=p_1)$ or $c^{(2)}=(c_1^+=p_2,~ c_2^+=p_1,~ c_1^-=c_2^-=0)$ (or both) solve (AUX).
\end{lemma}

\smallskip

\noindent \textbf{Proof of Lemma \ref{lemma-aux}:}\\
First, we claim that in any solution to (AUX) the constraints (\ref{eqn-bound1'-n=2}) and (\ref{eqn-bound2'-n=2}) are satisfied with equality. Indeed, suppose by contradiction that (\ref{eqn-bound1'-n=2}) holds with strict inequality at the optimum. If both $c_2^+<1$ and $c_2^->0$ then we can increase the former and decrease the latter while maintaining the equality in (\ref{eqn-n=2-equality}), contradicting the optimality. If $c_2^-=0$ then we must have $c_2^+<p_1$; equation (\ref{eqn-n=2-equality}) then implies that $c_1^+<p_2$, so that (\ref{eqn-bound2'-n=2}) is also strict; but then we can increase both $c_2^+$ and $c_1^+$ while maintaining all the constraints, which again contradicts the optimality. In the case $c_2^+=1$ we get a similar contradiction. This proves the claim for (\ref{eqn-bound1'-n=2}). The proof for (\ref{eqn-bound2'-n=2}) is analogous.

Second, since (\ref{eqn-bound1'-n=2}) and (\ref{eqn-bound2'-n=2}) are binding, we can express the objective and all the constraints using $c_1^+$ and $c_2^+$ only. After some simplifications this gives the following:
\begin{eqnarray*}
\max_{c_1^+, c_2^+}~  \frac{p_1(1-p_2)U_1^+ - p_2(1-p_1)U_1^-}{1-p_2}c_1^+  &+& \frac{p_2(1-p_1)U_2^+ - p_1(1-p_2)U_2^-}{1-p_1}c_2^+ + Const  \\
\textit{s.t. }~~~ \frac{p_1(1-p_2)+p_2(1-p_1)}{1-p_2}c_1^+ + \frac{(1-p_1)p_2^2}{1-p_2} &=& \frac{p_1(1-p_2)+p_2(1-p_1)}{1-p_1}c_2^+ + \frac{(1-p_2)p_1^2}{1-p_1}\\
p_2\le c_1^+ \le 1 ,~~ p_1\le c_2^+ \le 1. & &
\end{eqnarray*}
The feasible set of this problem is the interval in $\RR^2$ between the points $(c_1^+, c_2^+) = (1, 1)$ and $(c_1^+, c_2^+) = (p_2, p_1)$. Since the objective is affine, (at least) one of these corners is a maximizer. If the former is a maximizer then $c^{(1)}$ solves (AUX), while if the latter is a maximizer then $c^{(2)}$ does.
\hfill \qed

\bigskip

To prove the theorem, let $f$ be any anonymous and BIC SCF. Lemma \ref{lemma-n=2} implies that $c(f) = (c_1^+(f), c_1^-(f), c_2^+(f), c_2^-(f))$ satisfies inequalities (\ref{eqn-bound1'-n=2}) and (\ref{eqn-bound2'-n=2}). Moreover, $c(f)$ satisfies equation (\ref{eqn-n=2-equality}) since both sides are equal to $\EE(f(\tilde v_1, \tilde v_2))$. In addition, it is clear that each element of $c(f)$ is bounded between 0 and 1. Thus, $c(f)$ is feasible for program (AUX).

Now, the qualified majority rules $f^{(1)}$ and $f^{(2)}$ are clearly anonymous and BIC, and we have $c(f^{(1)}) = c^{(1)}$ and $c(f^{(2)}) = c^{(2)}$. Therefore,
$$W(f) = W'(c(f))\le \max\{W'(c(f^{(1)})),~ W'(c(f^{(2)}))\} = \max\{W(f^{(1)}),~ W(f^{(2)})\},$$
where the inequality follows from Lemma \ref{lemma-aux} and the two equalities from (\ref{eqn-welfare}). This completes the proof.
\hfill \qed


\subsection{Proof of Theorem \ref{theorem-n>3}}\label{subsec-proof2}

Fix $n\ge 3$. Let $M>0$ be sufficiently large so that the two inequalities 
\begin{equation}\label{eqn-unanimity-optimal}
    \frac{2}{n}\left[-M^2+M+n-2\right]+\frac{n-2}{n}\left[2M+n-4\right] < 0
\end{equation}
and 
\begin{equation}\label{eqn-improve-W}
    2M-(n-2) > 0
\end{equation}
are satisfied. We consider environments with $n$ agents and where $V=\{-M^2,-1,1,M\}$. Environments are indexed by $\epsilon\in[0,\bar \epsilon]$, for some small $\bar\epsilon>0$. The $\epsilon$ environment is denoted $\Gamma^\epsilon = (G_1^\epsilon,\ldots,G_n^\epsilon)$, where the value distributions are given by the following table:

\begin{center}
\begin{tabular}{c||c|c|c|c|}
$V$ & $-M^2$ & -1 & 1 & $M$ \\
\hline
$G^\epsilon_1,G^\epsilon_2$ & 0.5-$\epsilon$ & $\epsilon$ & $\epsilon$ & 0.5-$\epsilon$   \\ 
\hline
$G^\epsilon_3,\ldots,G_n^\epsilon$ &  $\epsilon$ & 0.5-$\epsilon$ & 0.5-$\epsilon$ & $\epsilon$ \\
\hline   
\end{tabular}
\end{center}

We refer to agents 1 and 2 as high-stakes agents and to the rest as low-stakes agents. The welfare of any SCF $f$ in $\Gamma^\epsilon$ is denoted $W^\epsilon(f)$. We break the proof into a sequence of lemmas similarly to the steps described in the main text.

\begin{lemma}\label{lemma-unanimity}
    For all sufficiently small $\epsilon>0$ the unanimity rule $f^{(n)}$ maximizes welfare in $\Gamma^\epsilon$ among all ordinal and anonymous SCFs.
\end{lemma}

\smallskip

\noindent \textbf{Proof of Lemma \ref{lemma-unanimity}:}\\
Recall that maximizing welfare among ordinal and anonymous SCFs amounts to finding the optimal qualified majority rule $f^{(k)}$, $k\in \{1,\ldots,n\}$. Suppose first that $\epsilon=0$. If all $n$ agents have positive values then setting $f=1$ is clearly optimal. Now, conditional on the event that $n-1$ agents have positive values and one agent has negative value, the probability that the negative value is of a high-stake agent is $\frac{2}{n}$. Thus, the expected welfare from choosing $R$ conditional on this event is 
$$\frac{2}{n}\left[-M^2+M+n-2\right]+\frac{n-2}{n}\left[2M+n-4\right].$$
But from (\ref{eqn-unanimity-optimal}) this expression is negative, so welfare is higher by setting $f=0$. This implies that $f^{(n)}$ yields strictly higher expected welfare than any other qualified majority rule in $\Gamma^0$, namely $W^0(f^{(n)})>W^0(f^{(k)})$ for any $k=1,\ldots,n-1$. Since $W^\epsilon(f)$ is continuous in $\epsilon$ for any fixed $f$, it follows that $W^\epsilon(f^{(n)})>W^\epsilon(f^{(k)})$ for any $k=1,\ldots,n-1$ and for any sufficiently small $\epsilon>0$.
\hfill \qed

\bigskip

\begin{lemma}\label{lemma-better-cardinal}
  In $\Gamma^0$ there exists a BIC and anonymous SCF $f^*$ that yields higher welfare than $f^{(n)}$. 
\end{lemma}

\smallskip

\noindent \textbf{Proof of Lemma \ref{lemma-better-cardinal}:}\\
Let $f^*$ be identical to  $f^{(n)}$, except that 
$$f^*(\{M,M,-1,\ldots,-1\}) = f^*(\{-M^2,1,\ldots,1\})=f^*(\{-M^2,-M^2,-M^2,1,\ldots,1\})=1.$$
To compare the welfare, note that the only positive probability profile at which $f^*$ and $f^{(n)}$ differ is $\{M,M,-1,\ldots,-1\},$ at which $f^*=1$ but $f^{(n)}=0$. Since $M$ was chosen to satisfy (\ref{eqn-improve-W}), it follows that $W^0(f^*)>W^0(f^{(n)})$. 

To check that $f^*$ is BIC, consider first agents 1. If this agent reports a negative value, either $-M^2$ or $-1$, then the probability that $R$ is chosen is zero; if they report a value of $1$ then the profiles of other agents' reports that have positive probability and lead to $R$ being selected are $\{M,1,\ldots,1\}$ and $\{-M^2,1,\ldots,1\}$, so the probability of $R$ is $\left(\frac{1}{2}\right)^{n-1}+\left(\frac{1}{2}\right)^{n-1}=\left(\frac{1}{2}\right)^{n-2}$; if the report is $M$ then the relevant profiles of others are $\{M,1,\ldots,1\}$ and $\{M,-1,\ldots,-1\}$, so the probability of $R$ is again $\left(\frac{1}{2}\right)^{n-1}+\left(\frac{1}{2}\right)^{n-1}=\left(\frac{1}{2}\right)^{n-2}$. Hence, agent 1 has no profitable deviation from truth-telling. The same calculations apply to agent 2. For agents 3 to $n$, whatever value they report there is a unique profile of reports of other agents that have positive probability and leads to $R$ being selected, so the probability of $R$ is independent of their own report and equals $\left(\frac{1}{2}\right)^{n-1}$. This shows that $f^*$ is BIC. 
\hfill \qed

\bigskip

\begin{lemma}\label{lemma-approximate}
  If $f$ is a BIC and anonymous SCF in $\Gamma^0$ then there exists a family of anonymous SCFs $\{f_\epsilon\}_{\epsilon\in(0,\bar\epsilon]}$ such that (i) each $f_\epsilon$ is BIC in $\Gamma_\epsilon$ and (ii) $\lim_{\epsilon\to 0}f_\epsilon = f$.
\end{lemma}

\smallskip

\noindent \textbf{Proof of Lemma \ref{lemma-approximate}:}\\
Consider the set $H^\epsilon$ of all functions $h:V^n\to\RR$ that are anonymous and satisfy the BIC equality constraints in $\Gamma^\epsilon$ (we will add the BIC inequality constraints and the restriction that the range is $[0,1]$ later). Since these are all linear homogeneous equality constraints, $H^\epsilon$ is the kernel of a linear map. The first step of the proof is to show that the rank of this map is constant for $\epsilon\in[0,\bar\epsilon]$. Denote by $r$ the rank of the matrix that defines only the anonymity constraints, which does not change with $\epsilon$. We claim that in $\Gamma^0$ the rank of matrix including all constraints is $r+4$. Indeed, since there are only two types of agents, an anonymous function $h$ satisfies the BIC equality constraints for all agents if and only if they hold for agents 1 and 3. Consider the four BIC equality constraints for these two agents:  
\begin{eqnarray}
    \EE[h(-M^2,\tilde v_{-1})] &=& \EE[h(-1,\tilde v_{-1})], \label{eqn-BIC-1-negative}\\
    \EE[h(M,\tilde v_{-1})] &=& \EE[h(1,\tilde v_{-1})],\label{eqn-BIC-1-positive}\\
    \EE[h(-M^2,\tilde v_{-3})] &=& \EE[h(-1,\tilde v_{-3})], \textit{ and }\label{eqn-BIC-3-negative}\\
    \EE[h(M,\tilde v_{-3})] &=& \EE[h(1,\tilde v_{-3})]. \label{eqn-BIC-3-positive}
\end{eqnarray}
We show that for each of the equalities (\ref{eqn-BIC-1-negative})-(\ref{eqn-BIC-3-positive}) there exists an anonymous $h$ that does not satisfy this constraint but does satisfy all other three constraints. For (\ref{eqn-BIC-1-negative}), take $h(\{-M^2,-1,\ldots,-1\})=1$ and $h(v)=0$ otherwise; then $\EE[h(-M^2,\tilde v_{-1})] = 0$, $\EE[h(-1,\tilde v_{-1})] = \left(\frac{1}{2}\right)^{n-1}$, and in all other constraints both sides are equal to zero; for (\ref{eqn-BIC-1-positive}) take $h(\{M,1,\ldots,1\})=1$ and $h(v)=0$ otherwise; for (\ref{eqn-BIC-3-negative}) take $h(\{-M^2,-M^2,-M^2,-1\ldots,-1\})=1$ and $h(v)=0$ otherwise; and for (\ref{eqn-BIC-3-positive}) take $h(\{M,M,M,1\ldots,1\})=1$ and $h(v)=0$ otherwise.    

Now, since the rank of a matrix is lower semi-continuous in its entries \citep[page 216]{meyer2010matrix}, and since these entries are continuous in $\epsilon$, it follows that for all sufficiently small $\epsilon>0$ the rank is at least $r+4$. However, the rank can't be strictly larger than $r+4$ since, for any $\epsilon$, if the anonymity constraints and (\ref{eqn-BIC-1-negative})-(\ref{eqn-BIC-3-positive}) hold then all other BIC equality constraints hold as well. This proves that the rank is constant on some non-trivial interval $[0,\bar \epsilon]$.

It now follows from e.g.\ \citet[Theorem 1]{weiss1969dolevzal} that $H^\epsilon$ is lower hemi-continuous at $\epsilon=0$. Namely, for every $f$ that is anonymous and BIC in $\Gamma^0$ there exists a family $\{h_\epsilon\}$ such that $h_\epsilon\in H^\epsilon$ for every $\epsilon$ and such that $h_\epsilon\to f$.

To finish the proof we need to show that lower hemi-continuity is preserved when one adds the BIC inequality constraints and the constraints that $0\le h\le 1$. For this purpose, denote by $f_{half}$ the constant SCF with $f(v)=0.5$ for all $v\in V^n$, and consider the SCF $\bar f:= 0.5 f^{(n)}+0.5f_{half}$. Note that, for every $\epsilon\in[0,\bar\epsilon]$, (i) $\bar f$ is BIC as a convex combination of two BIC SCFs, (ii) $\bar f(v)$ is strictly between zero and one for every $v$, and (iii) under $\bar f$ all BIC inequality constraints hold strictly, since reporting a positive value strictly increases the probability that $R$ is chosen in $f^{(n)}$. 

Fix some BIC and anonymous SCF $f$ in $\Gamma^0$. Let $\{h_\epsilon\}$ satisfy $h_\epsilon\in H^\epsilon$ and $\lim_\epsilon h_\epsilon=f$. For each $\epsilon$ define 
$$\alpha_\epsilon:=\min\{0\le \alpha\le 1 : \alpha \bar f +(1-\alpha)h_\epsilon \textit{ is BIC and bounded in } [0,1]\}.$$
Note that $\alpha_\epsilon$ is well-defined since the set of $\alpha$'s which satisfy both requirements is non-empty (contains $\alpha=1$) and closed. Furthermore, we claim that $\lim_{\epsilon\to 0}\alpha_\epsilon=0$. Indeed, if that was not the case then for some sequence $\epsilon_k\downarrow 0$ we would have $\alpha_{\epsilon_k}\to \alpha_0>0$. But $\alpha_0 \bar f +(1-\alpha_0)f$ satisfies all the inequality constraints strictly, contradicting the minimality of $\alpha_{\epsilon_k}$ for small enough $\epsilon_k$. Finally, by defining $f_\epsilon= \alpha_\epsilon \bar f +(1-\alpha_\epsilon)h_\epsilon$, we have that the family $\{f_\epsilon\}$ satisfies the requirements of the lemma.
\hfill\qed
 
\bigskip

We can now conclude the proof of the theorem. Let $f^*$ be as in Lemma \ref{lemma-better-cardinal} and let $\{f_\epsilon\}$ be as in Lemma \ref{lemma-approximate} and converge to $f^*$. Then
$$\lim_{\epsilon\to 0} W^\epsilon(f_\epsilon) = W^0(f^*)>W^0(f^{(n)})=\lim_{\epsilon\to 0} W^\epsilon(f^{(n)}),$$
where the first equality follows from $f_\epsilon\to f^*$ and continuity, the inequality is by the construction of $f^*$, and the last equality is again by continuity. It follows that for all $\epsilon>0$ small enough 
$$W^\epsilon(f_\epsilon)> W^\epsilon(f^{(n)}).$$
But from Lemma \ref{lemma-unanimity} $f^{(n)}$ is the optimal ordinal anonymous rule for small $\epsilon>0$, implying that an ordinal rule cannot be a solution to (\ref{eqn-max-welfare}).
\hfill\qed


\section{Code}\label{sec-code}

\begin{lstlisting}
# 0) Install Z3 (run once in an internet-enabled environment)
!pip install z3-solver

from z3 import Real, Optimize, sat
from fractions import Fraction
from itertools import product, combinations_with_replacement

# ====== Configuration: value set, agents, and distributions =======
V = [-100, -1, 1, 10]
agents = ["High", "High", "Low"]
# 3-agent setup, for 4 agents use e.g. ["High", "High" ,"Low" ,"Low"]

# Type-specific probability distributions
prob = {
"High": { 10:Fraction(499,1000), -100:Fraction(499,1000),  1:Fraction(1,1000), -1:Fraction(1,1000)},
"Low": { 1:Fraction(499,1000), -1:Fraction(499,1000),  10:Fraction(1,1000), -100:Fraction(1,1000)},
}
# ==============================================================

# 1) Print the probability distributions
print("=== Probability Distribution ===")
for t in sorted(prob):
    print(f"Type {t}:")
    for v,p in sorted(prob[t].items()):
        print(f"  v = {v:>2} → {p} (≈ {float(p):.6f})")
print()

# 2) Prepare Z3 optimizer
multisets = list(combinations_with_replacement(V, len(agents)))
f = {ms: Real(f"f_{'_'.join(map(str,ms))}") for ms in multisets}
opt = Optimize()

# Domain constraints: 0 ≤ f_ms ≤ 1
for var in f.values():
    opt.add(var >= 0, var <= 1)

def pr(profile):
    """Compute the probability of a type profile."""
    x = Fraction(1,1)
    for v,t in zip(profile, agents):
        x *= prob[t][v]
    return x

# 3) Objective: maximize expected welfare
opt.maximize(
    sum(pr(p) * sum(p) * f[tuple(sorted(p))]
        for p in product(V, repeat=len(agents)))
)

# 4) Incentive-compatibility constraints per agent type
for t in sorted(set(agents)):
    idx = agents.index(t)
    E = {v:0 for v in V}
    Pacc = {v:0 for v in V}

    # Build interim expected allocation and marginal probabilities
    for p in product(V, repeat=len(agents)):
        m   = tuple(sorted(p))
        prp = pr(p)
        Pacc[p[idx]] += prp
        E[p[idx]]   += prp * f[m]

    neg_vs = sorted(v for v in V if v < 0)
    pos_vs = sorted(v for v in V if v > 0)

    # Flatness among negative reports
    for v1,v2 in zip(neg_vs, neg_vs[1:]):
        opt.add(E[v1] * Pacc[v2] == E[v2] * Pacc[v1])

    # Flatness among positive reports
    for v1,v2 in zip(pos_vs, pos_vs[1:]):
        opt.add(E[v1] * Pacc[v2] == E[v2] * Pacc[v1])

    # Monotonicity: max negative ≤ min positive
    if neg_vs and pos_vs:
        opt.add(
            E[neg_vs[-1]] * Pacc[pos_vs[0]] <= 
            E[pos_vs[0]] * Pacc[neg_vs[-1]]
        )

# 5) Solve the optimization
if opt.check() == sat:
    model = opt.model()
    f_star = {
        ms: Fraction(model[f[ms]].numerator_as_long(),
                     model[f[ms]].denominator_as_long())
        for ms in multisets
    }
else:
    raise RuntimeError("No feasible solution found")

# 6) Print the nonzero entries of the optimal rule
print("=== Optimal IC rule f* (nonzero) ===")
for ms,val in sorted(f_star.items()):
    if val != 0:
        print(f"  {ms} → {val}  (≈ {float(val):.6f})")
print()

# 7) Compute and display expected welfare
w_opt = sum(pr(p) * sum(p) * f_star[tuple(sorted(p))]
            for p in product(V, repeat=len(agents)))
print(f"=== Expected welfare (IC rule): {w_opt} (≈ {float(w_opt):.6f}) ===\n")

# 8) Compare against QMR_k benchmarks
print("=== QMR_k Expected Welfare ===")
for k in range(len(agents) + 1):
    wq = sum(pr(p) * sum(p) * (1 if sum(v > 0 for v in p) >= k else 0)
             for p in product(V, repeat=len(agents)))
    print(f"  QMR_{k}: {wq} (≈ {float(wq):.6f})")

# ======================================================================
# ===            Compute utilitarian-optimal Weighted-Majority Rule  ===
# ======================================================================
# 1) Calculate individual weights w_i and the quorum q
weights = []                    # list of Fraction
q       = Fraction(0, 1)        # reform is chosen iff summed weights > q
for t in agents:
    wr = ws = Fraction(0, 1)
    pr_pos = pr_neg = Fraction(0, 1)
    for v, pv in prob[t].items():
        if v > 0:                       # types that prefer reform
            wr     += pv * v            # u_r − u_s = v
            pr_pos += pv
        elif v < 0:                     # types that prefer status-quo
            ws     += pv * (-v)         # u_s − u_r = −v
            pr_neg += pv
    wr_tilde = wr / pr_pos if pr_pos else Fraction(0, 1)
    ws_tilde = ws / pr_neg if pr_neg else Fraction(0, 1)
    weights.append(wr_tilde + ws_tilde)
    q += ws_tilde                       # quorum is the sum of status-quo losses

print("\n=== Optimal WMR parameters ===")
for i, (w, t) in enumerate(zip(weights, agents), 1):
    print(f"  Agent {i} (type {t}): weight = {w} (≈ {float(w):.6f})")
print(f"  Quorum q = {q} (≈ {float(q):.6f})")

# 2) Define the WMR decision rule
def wmr_rule(profile):
    w_support = sum(w for w, v in zip(weights, profile) if v > 0)
    if w_support > q:  return Fraction(1, 1)   # choose reform
    if w_support < q:  return Fraction(0, 1)   # choose status-quo
    return Fraction(1, 2)                      # tie → ½

# 3) Expected welfare of the optimal WMR
w_wmr = sum(pr(p) * sum(p) * wmr_rule(p) for p in product(V, repeat=len(agents)))
print(f"\n=== Expected welfare (optimal WMR): {w_wmr} "
      f"(≈ {float(w_wmr):.6f}) ===\n")

# ======================================================================
# ===     Find the welfare-maximizing QMR_k and compare all three     ==
# ======================================================================
w_qmr = {}
for k in range(len(agents) + 1):
    wq = sum(pr(p) * sum(p) * (1 if sum(v > 0 for v in p) >= k else 0)
             for p in product(V, repeat=len(agents)))
    w_qmr[k] = wq
k_star       = max(w_qmr, key=w_qmr.get)
w_qmr_best   = w_qmr[k_star]

# 4) Relative performance (percent of WMR welfare)
pct_qmr = float(w_qmr_best) / float(w_wmr) * 100
pct_ic  = float(w_opt)      / float(w_wmr) * 100

print("=== Welfare ratios relative to optimal WMR ===")
print(f"  • Best QMR (k = {k_star}) welfare: {w_qmr_best} "
      f"(≈ {float(w_qmr_best):.6f})  → {pct_qmr:.2f}% of WMR")
print(f"  • Anonymous IC-optimal cardinal rule welfare: {w_opt} "
      f"(≈ {float(w_opt):.6f})  → {pct_ic:.2f}% of WMR")
print("==============================================================")
\end{lstlisting}


\end{document}